\documentclass[aps,prd,twocolumn,showpacs,groupedaddress,preprintnumbers,superscriptaddress,amsmath,amssymb,floatfix,longbibliography]{revtex4-1}

\usepackage[colorlinks=true, filecolor=blue, citecolor=blue,urlcolor=blue]{hyperref}
\usepackage{url}

\usepackage{bm,pifont}
\usepackage{amssymb}
\usepackage{amsmath}
\usepackage{tabularx}
\usepackage{amsfonts}
\usepackage{multirow}
\usepackage{tabularx}
\usepackage{booktabs}
\usepackage{epsfig}
\usepackage{graphicx}
\usepackage{multirow}
\usepackage{tabularx}
\usepackage{color}
\usepackage{slashed}
\usepackage{orcidlink}
\newcommand{\seq}{\begin{subequations}}
\newcommand{\sen}{\end{subequations}}
\newcommand{\eq}{\begin{eqnarray}}
\newcommand{\en}{\end{eqnarray}}

\def\shiftdown#1{#1\llap{\lower.04ex\hbox{#1}}}

\def\nn{\nonumber}

\begin{document}

\title{Constraints on dark axion portal: missing energy and fermion EDMs} 


\author{Sergei N.~Gninenko \orcidlink{0000-0001-6495-7619}} 
\affiliation{Institute for Nuclear Research, 117312 Moscow, Russia}
\affiliation{Bogoliubov Laboratory of Theoretical Physics, JINR, 141980 Dubna, Russia} 
\affiliation{Millennium Institute for Subatomic Physics at
the High-Energy Frontier (SAPHIR) of ANID, \\
Fern\'andez Concha 700, Santiago, Chile}

\author{ N.~V.~Krasnikov \orcidlink{0000-0002-8717-6492}}
\affiliation{Institute for Nuclear Research, 117312 Moscow, Russia}
\affiliation{Bogoliubov Laboratory of Theoretical Physics, JINR, 141980 Dubna, Russia} 

\author{Valery~E.~Lyubovitskij \orcidlink{0000-0001-7467-572X}}
\affiliation{Institut f\"ur Theoretische Physik, Universit\"at T\"ubingen, \\
Kepler Center for Astro and Particle Physics, \\ 
Auf der Morgenstelle 14, D-72076 T\"ubingen, Germany} 
\affiliation{Millennium Institute for Subatomic Physics at
the High-Energy Frontier (SAPHIR) of ANID, \\
Fern\'andez Concha 700, Santiago, Chile}

\author{Sergey~Kuleshov~\orcidlink{0000-0002-3065-326X}}
\affiliation{Millennium Institute for Subatomic Physics at
the High-Energy Frontier (SAPHIR) of ANID, \\
Fern\'andez Concha 700, Santiago, Chile}
\affiliation{Center for Theoretical and Experimental Particle Physics,
Facultad de Ciencias Exactas, Universidad Andres Bello,
Fernandez Concha 700, Santiago, Chile}

\author{Alexey~S.~Zhevlakov \orcidlink{0000-0002-7775-5917}} 
\affiliation{Bogoliubov Laboratory of Theoretical Physics, JINR, 141980 Dubna, Russia} 
\affiliation{Matrosov Institute for System Dynamics and 
	Control Theory SB RAS, \\  Lermontov str., 134, 664033, Irkutsk, Russia }

\author{I.~V.~Voronchikhin \orcidlink{0000-0003-3037-636X}}
\affiliation{Institute for Nuclear Research, 117312 Moscow, Russia}
\affiliation{ Tomsk Polytechnic University, 634050 Tomsk, Russia}

\author{D.~V.~Kirpichnikov \orcidlink{0000-0002-7177-077X}}
\email[\textbf{e-mail}: ]{dmbrick@gmail.com}
\affiliation{Institute for Nuclear Research, 117312 Moscow, Russia}

\begin{abstract}
We study a model in which a new interactions between the Standard Model (SM) photon and both the dark photon ($\gamma_D)$ and an ALP ($a$) are described  by the dark axion portal operator. The implications of this dark axion portal scenario for electron fixed-target experiments are presented. 
In particular, we investigate the missing energy signatures associated with production of dark photons and their subsequent invisible decays into stable dark sector fermions, 
$\gamma_D \to \chi \bar{\chi}$. We discuss the discovery potential for such a scenario and derive projected sensitivity curves for the NA64$e$ and LDMX experiments.
Furthermore, novel constraints of the NA64$e$ for 
$9.37\times 10^{11}$ electrons on target are derived by considering two production mechanisms for invisible states: (i)~the bremsstrahlung-like emission of an 
$a\gamma_D$ pair, $e N \to e N \gamma^* (\to a \gamma_D)$, and (ii)~the exclusive vector meson photoproduction, 
$\gamma^* N \to N V$, followed by the invisible decays 
of vector mesons, $V \to a \gamma_D$. Additionally, the constraints on the parameter space of $CP$-violating, 
fermion-specific ALP and dark photon couplings are established. These constraints are derived from current experimental bounds 
on the electric dipole moments (EDMs) of SM fermions, 
incorporating loop-induced contributions to the EDMs of the electron, muon, and neutron. 
\end{abstract}

\maketitle

\section{Introduction}

Axion-like particles (ALP or $a$) represent a well-motivated class of pseudo-scalar bosons that emerge naturally in numerous extensions of the Standard Model (SM). Originally posited as a dynamical solution to the strong CP problem, their theoretical foundation has been expanded significantly, as reviewed 
in~\cite{Peccei:1977ur,DiLuzio:2020wdo}. Beyond this primary role, ALPs provide a viable framework for addressing  phenomenological puzzle, serving as viable candidates for the cosmological dark matter (DM) relic 
density~\cite{Boehm:2003hm,Dolan:2014ska,Hochberg:2018rjs}.

The phenomenological landscape of ALPs extends to more exotic scenarios, such as those featuring lepton flavor violation. 
Such models, where ALPs mediate transitions between different lepton generations, have been the subject of detailed 
theoretical and experimental investigation~\cite{Han:2020dwo,Davoudiasl:2021mjy,Gninenko:2022ttd,Bauer:2020jbp}. Recent reviews and a multitude of studies, spanning beam-dump experiments, fixed-target facilities, colliders, RF cavities, and astrophysical probes, explore  
various signatures and constraints on these 
particles~\cite{Choi:2020rgn,Dusaev:2020gxi,NA64:2020qwq,Ishida:2020oxl,Sakaki:2020mqb,Brdar:2020dpr,Salnikov:2020urr,Salnikov:2024shh,Kahn:2022uko,Darme:2020sjf,Dev:2021ofc,Abramowicz:2021zja,Fortin:2021cog,Asai:2021ehn,Blinov:2021say,PrimEx:2010fvg,Gninenko:2016kpg,NA64:2016oww,Gninenko:2017yus,Gninenko:2019qiv,Banerjee:2019pds,Andreev:2021fzd,NA64:2021xzo,Blinov:2020epi,Beattie:2018xsk,Pankratov:2025cby,Jodlowski:2024ayf}.

Recent work has introduced a novel interaction framework known 
as the dark axion portal~\cite{Kaneta:2016wvf,Kaneta:2017wfh}. This effective operator facilitates the production of dark 
photons in the early Universe and, for sufficiently low-mass dark photon states, provides a viable mechanism for generating the observed cosmological dark matter relic density. The portal 
is characterized by vertices coupling the axion (or a generic ALP) simultaneously to the SM photon and the dark photon, $a$-$\gamma$-$\gamma_D$.

Building upon the theoretical framework proposed in 
Refs.~\cite{deNiverville:2018hrc,deNiverville:2019xsx,Zhevlakov:2022vio}, this work investigates the phenomenology 
of the dark axion portal in a scenario where the associated dark photon decays predominantly into 
dark sector (DS) states. In this model, the dark photon is identified as the gauge boson of a 
hidden $U_D(1)$ symmetry, thereby acting as the primary mediator between the Standard Model and 
DS particles through the portal interaction.

We demonstrate that this scenario yields a rich phenomenology accessible through missing energy signatures at 
electron beam fixed-target experiments. Specifically, we conduct an analysis of the discovery potential for such 
signals at the current experiment NA64$e$ \cite{NA64:2025ddk,Gninenko:2017yus,Gninenko:2019qiv,Banerjee:2019pds,NA64:2021xzo,Andreev:2021fzd,NA64:2023wbi} and the proposed LDMX facility \cite{Mans:2017vej,Berlin:2018bsc,LDMX:2018cma,Ankowski:2019mfd,Schuster:2021mlr,Akesson:2022vza,LDMX:2025bog}.

We also develop the ideas presented in 
Refs.~\cite{Kirpichnikov:2020tcf,Kirpichnikov:2020lws} where
the implication of light sub--GeV bosons for  electric dipole moments (EDM) of 
fermions and $CP$--odd dark axion portal coupling was discussed in detail. 
In  particular, we argue that the EDM of 
fermions  can be induced by: (i) $CP$--odd  Yukawa--like  couplings of ALP,  (ii) 
$CP$--even  interaction of SM fermions and dark photon and (iii) $CP$--even dark 
axion  portal coupling. As a result, one can obtain  bounds on the  corresponding  couplings. 

This work is structured as follows. Sec.~\ref{DescriptionSection} details the theoretical framework and benchmark scenarios for the dark axion 
portal. Sec.~\ref{ExperimentalBenchmark} outlines the missing energy signatures relevant for dark matter production in fixed-target experiments.
In Sec.~\ref{Sec:Experimenta} we outline benchmark experiments.  
In Sec.~\ref{Sec:Sensitivity}, we derive constraints on the portal's effective couplings using data from electron-beam fixed-target facilities. 
Sec.~\ref{EDMsectionLabel} explores the generation of electric dipole moments for Standard Model fermions within dark axion portal framework 
incorporating  additional $CP$-violating couplings. Finally, Section~\ref{ResultsSection} provides a summary  and conclusions. 

\section{The description of the benchmark scenarios
\label{DescriptionSection}}
 
The coupling between dark and electromagnetic photons can
be associated with ALPs through specific interaction~\cite{Kaneta:2016wvf,Kaneta:2017wfh}. 
The effective Lagrangian for such nonrenormalizable dark axion portal takes the form
\begin{equation}
\label{LDAP}
\mathcal{L}_{\mbox{\scriptsize dark\! axion\! portal}} \supset \frac{g_{a \gamma \gamma_D}}{2} a F_{\mu\nu} \widetilde{F}^{\prime\mu\nu}\,,  
\end{equation}
that is the 
interaction  between ALP and both SM and dark photon, 
where $F^{\mu\nu}$ and $\widetilde{F}^{\prime\mu\nu}$ 
are the stress tensors of the visible and dark photons, 
respectively, $g_{a \gamma \gamma_D}$ is the coupling constant. 

This work explores a benchmark extension of the dark axion portal framework by incorporating a minimal dark sector candidate. Within this model, the dark photon, emerging from a hidden $U_D(1)$ gauge symmetry, acts as the mediator between the  Standard Model sector and a dark fermionic sector. The relevant Lagrangian is given by:
\begin{equation}
\mathcal{L}_{ { \rm \scriptsize DS}} \supset  \bar{\chi} \left( i \gamma^\mu D_\mu - m_{\chi} \right) \chi,
\label{MinimalSetupCoupling1}
\end{equation}
 where the DS field $\chi$  is a 
Dirac fermion representing a dark sector particle with mass $m_\chi$. Its covariant derivative is 
$D_\mu = \partial_\mu + i g_D A'_\mu$, introducing the gauge coupling $g_D$ associated with the hidden 
$U_D(1)$ charge. A key phenomenological assumption of this setup is that the dark photon decays predominantly 
into a pair of these dark fermions, $A' \to \chi \bar{\chi}$, resulting in an invisible final state thus making it interesting for searching, in particular, 
in fixed target experiments, such as NA64e and LDMX. The phenomenological consequences 
of these signatures are analyzed in Sec.~\ref{DescriptionSection}, 
\ref{ExperimentalBenchmark},~\ref{Sec:Experimenta}, 
and ~\ref{Sec:Sensitivity}. 

One can assume the existence of the
 the Yukawa--like couplings of the  ALP with SM 
 fermions~\cite{Stadnik:2017hpa,Dzuba:2018anu,Flambaum:2009mz,Kirpichnikov:2020lws,Zhevlakov:2022vio}, 
 \begin{equation}
 \mathcal{L}_{\!  \not CP} \supset \sum_{f=e,\mu,n}   g^{a}_f \, a \bar{f} f \label{CPoddALPcoupling}
 \end{equation}
 violating the  $CP$ symmetry,
thus they can induce the  electric dipole moment (EDM) of SM 
fermions.  It is worth to estimate the bounds on the combination of couplings from the EDM of fermions in the 
framework of their additional model independent benchmark interactions with dark photon 
\begin{equation}
\mathcal{L} \supset \sum_{f=e,\mu,n}   e \epsilon_f \, A_\mu' \bar{f} \gamma^\mu f,
\label{ProtophobicVectorCoupling}
\end{equation}
where protophobic coouplings $e\epsilon_f$ of massive  hidden vector can be relevant to the exotic 
fifth force  scenarios addressing in Ref.~\cite{Feng:2016jff}.

The study of EDM fermion bounds, incorporating the Lagrangians in Eqs.~(\ref{LDAP}), 
(\ref{CPoddALPcoupling}), and (\ref{ProtophobicVectorCoupling}), is of particular interest in the 
present paper. In Sec.~\ref{EDMsectionLabel}, we discuss their phenomenological implications. 

 \begin{figure*}[!t]
\centering
\includegraphics[width=0.48\textwidth]{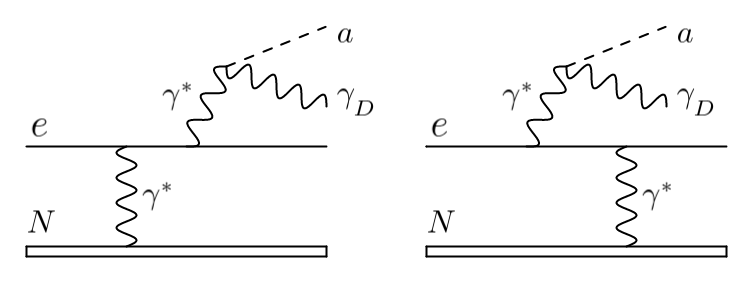}
\includegraphics[width=0.25\textwidth]{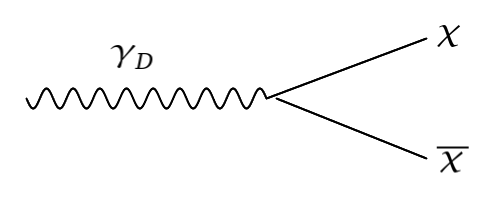}
\caption{ Feynman diagrams illustrating the bremsstrahlung-like production of an 
ALP and a dark photon in electron-nucleus scattering, 
\( e N \to e N a \gamma_D \). These diagrams correspond to the missing-energy 
signature within the minimal dark axion portal scenario, governed by the 
effective Lagrangian \(\mathcal{L} \supset \frac{1}{2} g_{a\gamma \gamma_D} a F_{\mu \nu} \widetilde{F}^{\prime \mu \nu}\). In this analysis, the produced dark 
photon decays predominantly into invisible dark sector fermions, \( \gamma_D \to \chi \bar{\chi} \). Diagrams involving photon emission from the nucleus line are 
omitted, as their contribution is suppressed by a factor of order \( (Z m_e / M_N)^2 \) for electron-nucleus scattering and is therefore negligible.
\label{MinimalALPportalBremsFeynman}}
\end{figure*}

\section{The missing energy signal
\label{ExperimentalBenchmark}} 

This section outlines the missing energy/momentum signatures in the $2\to4$  reactions which can be searched for  with the electron fixed-target experiments NA64$e$~\cite{Gninenko:2017yus,Gninenko:2019qiv,Banerjee:2019pds,NA64:2021xzo,Andreev:2021fzd} and LDMX~\cite{Mans:2017vej,Berlin:2018bsc,LDMX:2018cma,Ankowski:2019mfd,Schuster:2021mlr,Akesson:2022vza}. Specifically, the signature of interest arises from a reaction of a high-energy electron scattering on a heavy target nucleus $N$:
\begin{equation}
e N \to e N a \gamma_D ,
\label{generalMissinEnergyProcess}
\end{equation}
where the scattered  electron in the final state is accompanied by  a significant missing energy which is carried away by the ALP, $a$, and dark photon, $\gamma_D$ 
escaping detection.

Within the minimal dark axion portal scenario (\ref{LDAP}), one of the source of the   missing energy ($E_{\rm miss}$) is the bremsstrahlung-like production of an $a\gamma_D$ pair via an off-shell photon. The complete process,
$e N \to e N \gamma^* \to e N a \gamma_D$, is followed by the prompt invisible decay of the dark photon, $\gamma_D \to \bar{\chi}\chi$, into stable dark sector fermions. A representative Feynman diagram for this mechanism is shown in Fig.~\ref{MinimalALPportalBremsFeynman}.

While electron bremsstrahlung remains a primary mechanism for producing hidden 
sector particles, Ref.~\cite{Schuster:2021mlr} proposes a complementary source of 
missing-energy signatures (see e.~g.~Refs.~\cite{Banerjee:2025ejz,Gninenko:2023rbf,Zhevlakov:2023wel,Zhevlakov:2025dbh} for recent study  of meson decays to DM particles at fixed-target experiments). This alternative channel implies the exclusive 
photoproduction of a high-energy vector meson $V$ (such as $\rho$, $\omega$, 
$\phi$, or $J/\psi$) on a target nucleus $N$, via the reaction 
$\gamma^* N \to N V$. The initial off-shell photon $\gamma^*$ originates from 
standard bremsstrahlung, $e N \to e N \gamma^*$. 

\begin{figure*}[t!]
\centering
\includegraphics[width=0.33\textwidth]{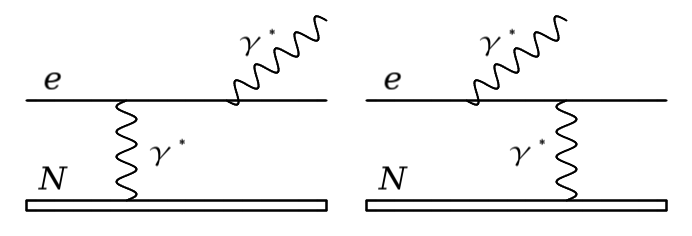}
\includegraphics[width=0.4\textwidth]{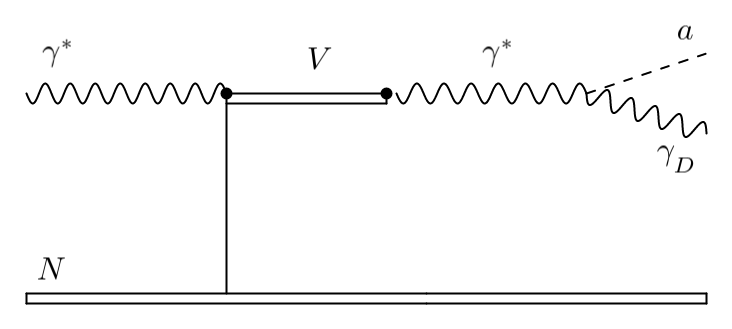}
\includegraphics[width=0.25\textwidth]{feynman_GammaD_To_chi_chi.png}
\caption{Feynman diagrams for the radiative invisible vector meson decay. A hard photon $\gamma^*$ is produced in the target, $eN\to eN \gamma^*$, and converts to a vector meson $V$ in the exclusive photoproduction process $\gamma^* N \to N V$ in the 
calorimeter. Then vector meson decays invisibly to 
the $a\gamma_D $ pair via mixing with the ordinary of-shell photon, $\gamma^*$. The produced dark photon dominantly decays into invisible DS particles, $\gamma_D\to \chi \bar{\chi}$. 
\label{JpsiFeynam}}
\end{figure*}

In this analysis, we focus on a specific invisible decay mode of these vector mesons. We consider the process, see Fig.~\ref{JpsiFeynam}, where a meson decays into an off-shell photon, which then converts into an ALP and a dark photon: $V \to \gamma^* \to a \gamma_D$ . This channel provides a distinct signature of the dark axion portal, contributing to the total missing energy in the final state.

In the dark axion portal scenario, the dark photon can decay via several 
channels. For masses satisfying $m_{\gamma_D} \gtrsim m_a$, 
a visible two-body decay into ALP and SM photon becomes kinematically allowed. Its partial decay width is given 
by~\cite{deNiverville:2019xsx}:
\begin{equation}
\Gamma_{\gamma_D \to a \gamma} = \frac{g_{a\gamma\gamma_D}^2}{96 \pi}
m_{\gamma_D}^3 \left(1-\frac{ m_a^2}{m_{\gamma_D}^2}\right)^3 .
\end{equation}

Simultaneously, the Lagrangian of  Eq.~(\ref{MinimalSetupCoupling1}) 
permits an invisible decay into a pair of dark sector fermions, 
$\gamma_D \to \bar{\chi}\chi$, provided $m_{\gamma_D} \gtrsim 2 m_\chi$. The corresponding partial width is~\cite{Bondi:2021nfp}:
\begin{equation}
\Gamma_{\gamma_D \to \bar{\chi}\chi} = \frac{g_D^2}{12 \pi} m_{\gamma_D}
\left(1+\frac{2 m_\chi^2}{m_{\gamma_D}^2}\right)
\left(1-\frac{4 m_\chi^2}{m_{\gamma_D}^2}\right)^{1/2}.
\end{equation}

This work focuses on the scenario where the dark photon decays invisibly into a dark fermion pair, $\gamma_D \to \bar{\chi}\chi$. We assume a branching fraction 
$\mbox{Br}_{\gamma_D \to \bar{\chi}\chi} \simeq 1$, which is naturally satisfied for $m_\chi \ll m_{\gamma_D}$. This condition requires the invisible decay width to dominate over the visible channel, 
$\Gamma_{\gamma_D \to \bar{\chi}\chi} \gg \Gamma_{\gamma_D \to a \gamma}$. Consequently, a hierarchy between the couplings is implied, $g_D \gg g_{a\gamma\gamma_D} m_{\gamma_{D}}$. Under this hierarchy, any produced dark photon decays predominantly into the dark sector.

Furthermore, to ensure the invisibility of the final state within the fixed-target detector, we require the ALP mass to be significantly smaller than the dark photon mass, $m_a \ll m_{\gamma_D}$. This kinematically suppresses the potentially visible decay $a \to \gamma \gamma_D$, guaranteeing that both the ALP and the dark sector fermions escape the experiment undetected, resulting in a clean missing-energy signature.

\subsection{Bremstrahlung-like production: $\gamma^* \to a \gamma_D$}

The expected number of missing-energy events from the process 
\( e N \to e N \gamma^* \to e N (a \gamma_D) \), where the  electron scatters off a target nucleus \( N \), can be estimated as:
\begin{equation}
N_{\gamma^* \to a\gamma_D} \simeq \mbox{EOT} \cdot \frac{\rho N_A}{A} L_T \int\limits_{E_{\rm min}}^{E_{\rm max}} dE_{\rm miss} \, \frac{d\sigma_{2\to4}(E_e)}{dE_{\rm miss}}.
\label{NumberOfMissingEv1}
\end{equation}
In this expression, \( E_e \) is the initial electron beam energy, \( \mbox{EOT} \) is the total number of electrons on target, and \( \rho \), \( A \), and \( L_T \) denote the density, atomic mass, and effective length of the target, respectively. The symbol \( N_A \) represents Avogadro’s number. The differential cross section \( d\sigma_{2\to4}(E_e)/dE_{\rm miss} \) describes the \( 2\to4 \) bremsstrahlung-like process \( e N \to e N \gamma^* \to e N a \gamma_D  \), and \( E_{\rm miss} = E_a + E_{\gamma_D} \) is the total energy carried away by the invisible ALP and dark photon system. The integration limits \( E_{\rm min} \) and \( E_{\rm max} \) are set by the specific energy acceptance and selection criteria of the experimental setup.
 
We compute the relevant $2\to4$ cross sections using the CalcHEP software package~\cite{Belyaev:2012qa}. To this end, we have extended the SM implementation in CalcHEP by introducing two new massive particles: the ALP $a$ and the dark photon $\gamma_D$. Their interaction with the SM photon is incorporated via the effective Lagrangian term (\ref{LDAP}).
The target nucleus is modeled as a spin-1/2 particle with mass $M_N$, atomic mass $A$, and charge $Z$. Its coupling to the photon is implemented through the vertex factor $i e Z F(t) \gamma_\mu$, where $t = -q^2 > 0$ is the squared momentum transfer to the nucleus and $F(t)$ denotes the nuclear elastic form factor. Following~\cite{Bjorken:2009mm,Tsai:1986tx}, we employ the parameterization
\begin{equation}
F(t) = \frac{a^2 t}{1 + a^2 t} \, \frac{1}{1 + t / d},
\label{FFdefinitio1}
\end{equation}
with screening and finite-size parameters given by $a = 111 Z^{-1/3}/m_e$ 
and $d = 0.164 A^{-2/3}\,\mbox{GeV}^2$. This form factor is incorporated into the 
analytic expression for the squared matrix element 
$|\mathcal{M}_{eN\to eNa\gamma_D}|^2$ generated by CalcHEP, 
and the corresponding C++ source files have been modified accordingly.

Using the experimental parameters detailed in Sec.~\ref{Sec:Experimenta}, we numerically integrate the exact tree-level amplitude squared for the process
\( e N \to e N \gamma^*(\to a\gamma_D) \) over the full final-state phase space. 
This computation is performed within the CalcHEP framework. For a given target material and a fixed electron beam energy \( E_e \), we evaluate the total cross section \( \sigma_{\rm tot} \) as a function of the dark photon mass \( m_{\gamma_D} \), scanning the range \( 1 \, \text{MeV} \lesssim m_{\gamma_D} 
\lesssim 1 \, \text{GeV} \) while fixing the ALP mass at 
\( m_a \simeq 10 \, \text{keV} \).

The multidimensional phase-space integration is carried out using the VEGAS 
adaptive Monte Carlo algorithm. We configure the integration with 
\( N_{\rm session} = 10 \) independent runs, each employing 
\( N_{\rm calls} = 10^6 \) sampling points. The adaptive grid optimization within 
VEGAS achieves a numerical precision of approximately \( \mathcal{O}(0.1)\% \) to 
\( \mathcal{O}(0.01)\% \) in the final cross-section results.

\subsection{Production via vector meson decays: $V\to a \gamma_D$}

The expected number of missing-energy events from the process \( e N \to e N \gamma^* \to e N (a \gamma_D) \), where the  electron scatters off a target nucleus \( N \), can be estimated as:
\[
N_V = \mbox{EOT} \times f_{\rm brem} \times P_V,
\]
where EOT is the total number of electrons on target, \(f_{\rm brem}\) denotes the fraction of incident electrons that emit a hard bremsstrahlung photon capable of producing the meson, and \(P_V\) represents the probability for such a photon to undergo an exclusive photoproduction interaction on the target. Following the treatment in Ref.~\cite{Schuster:2021mlr}, the photoproduction probability can be approximated as
\[
P_V \simeq \frac{9}{7} \frac{\sigma_0^V X_0 f_{\rm nuc}^V}{m_p},
\label{PVexpression}
\]
with \(X_0\) the radiation length of the target material, \(\sigma_0^V\) the exclusive photoproduction cross section on a single nucleon, and \(f_{\text{nuc}}^V\) an \(\mathcal{O}(1)\) nuclear correction factor.

For the LDMX Phase-II design, Ref.~\cite{Schuster:2021mlr} estimates \(f_{\rm brem}^{\rm LDMX} \simeq 0.03\). Assuming the ultimate planned integrated luminosity of \(\mbox{EOT} \simeq 10^{16}\), the resulting yields for light vector mesons are
\begin{equation}
    N_\rho \simeq 3\times 10^{10}, \quad N_\omega \simeq 3\times 10^{9}, \quad N_\phi \simeq 5\times 10^{8}.
    \label{NimberOfMesonsLDMXultimate}
\end{equation}
These yields correspond to a total meson production ranging from \(\mathcal{O}(10^8)\) to \(\mathcal{O}(10^{10})\) events.

For the NA64e planned statistics of 
$\mbox{EOT}\simeq 5\times 10^{12}$ the typical numbers of vector mesons are estimated to be~\cite{Schuster:2021mlr} 
\begin{align}
 &  N_\rho \simeq 1.2\times 10^{8}, \,\, N_\omega \simeq 9\times 10^{6}, \label{NimberOfMesonsNA64ultimate}\\
 & N_\phi \simeq 8\times 10^{6}, \,\, 
    N_{J/\psi}\simeq 1.1\times 10^5,   \nn 
\end{align}
these values imply the typical fraction of hard bremsstrahlung 
electrons at the level of 
$f^{\rm NA64e}_{\rm brem}\simeq 0.5$. Remarkably, at LDMX 
energies, $J/\psi$ production is kinematically forbidden. 
However, its coherent $\gamma^* N \to  N J/\psi$ production 
is kinematically allowed at the NA64e facility.  

As a result, the typical signal yield associated with vector meson invisible decay into pair of ALP-dark photon particles  reads
\begin{equation}
    N_{V\to a \gamma_D} \simeq N_{V} \times \mbox{Br}_{V\to a \gamma_D},
    \label{NumbSignEVMesDec}
\end{equation}
where $N_{V}$ is the total number of the specific vector meson, 
$\mbox{Br}_{V\to a \gamma_D}$ is the invisible branching fraction of vector meson associated with dark ALP portal scenario. 
Specifically, it is given by
\begin{equation}
  \mbox{Br}_{V\to a \gamma_D} =  \Gamma_{V\to  a \gamma_D}/\Gamma^{V}_{\rm tot}  
\end{equation}
where $\Gamma_{\rm tot}^{V}$ and $\Gamma_{V\to  a \gamma_D}$  are the total and invisible decay width of the vector meson, respectively. In order to calculate 
$\Gamma_{V\to  a \gamma_D}$, we employ the vector meson
dominance (VMD) framework~\cite{OConnell:1995nse}. The relevant mixing between ordinary SM photon, $A_\mu$, and vector meson, $V_\mu$, reads
\begin{equation}
    \mathcal{L}_{\rm VMD} \supset  -\sum_{V} \frac{e \, m_{V}^2}{g_V} A_\mu
    V^\mu, \label{VMDmixing}
\end{equation}
where $m_V$ is a vector meson mass, $e$ is the elementary electric charge, and $g_V$ is a dimensionless coupling constant that can be extracted from the data on strong decays~\cite{ParticleDataGroup:2024cfk}. 
In Tab.~\ref{tab:VecMeSTypTerms} we collect all vector meson  terms relevant for our analysis.  

The amplitude for an on-shell vector meson $V$ to decay to $a \gamma_D$ via virtual 
photon $\gamma^*$ reads 
\begin{equation}
    \mathcal{M}_{V\to a \gamma_D} =    \epsilon^\mu(q) \frac{e m_V^2}{g_V}\frac{g_{\mu\nu}}{q^2} 
     g_{a\gamma\gamma_D} 
    \epsilon^{\nu \lambda \rho \sigma} k_\lambda q_\rho  \epsilon^{*}_\sigma(k), 
\end{equation}
where $q^\rho$ and $k^\lambda$ are the momenta of the vector meson and dark photon, respectively. Squaring and summing over final dark photon polarizations $\epsilon^{*}_\sigma(k)$ and averaging over initial meson polarizations $\epsilon_\mu(q)$ yields the decay rate in the rest frame of the meson: 
\begin{equation}
\Gamma_{V\to a \gamma_D} = \frac{e^2 g_{a\gamma \gamma_D}^2}{96\pi} 
\frac{m_V^3}{g_V^2}\left(1-\frac{m_{\gamma_D}^2}{m_V^2} 
\right)^3 \,.
\label{VinvisDecW}
\end{equation}
To evaluate the decay width (\ref{VinvisDecW}), we employ the FeynCalc package~\cite{Shtabovenko:2016sxi} within the Wolfram Mathematica environment~\cite{Mathematica}. This calculation assumes the mass hierarchy \( m_{\gamma_D} \gg m_a \), allowing us to safely neglect the ALP mass, consistent with the assumptions outlined earlier.

\begin{figure*}[!tbh]
\centering
\includegraphics[width=0.495\textwidth]{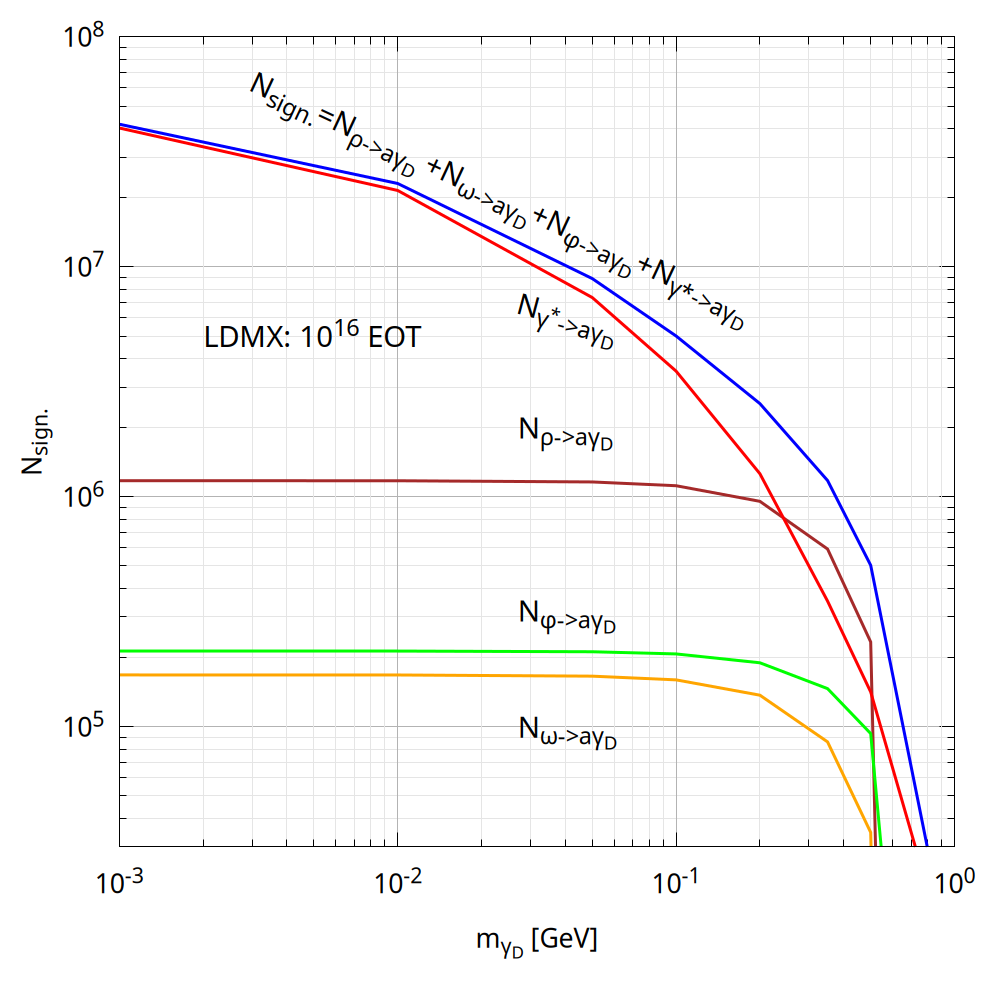}
\includegraphics[width=0.495\textwidth]{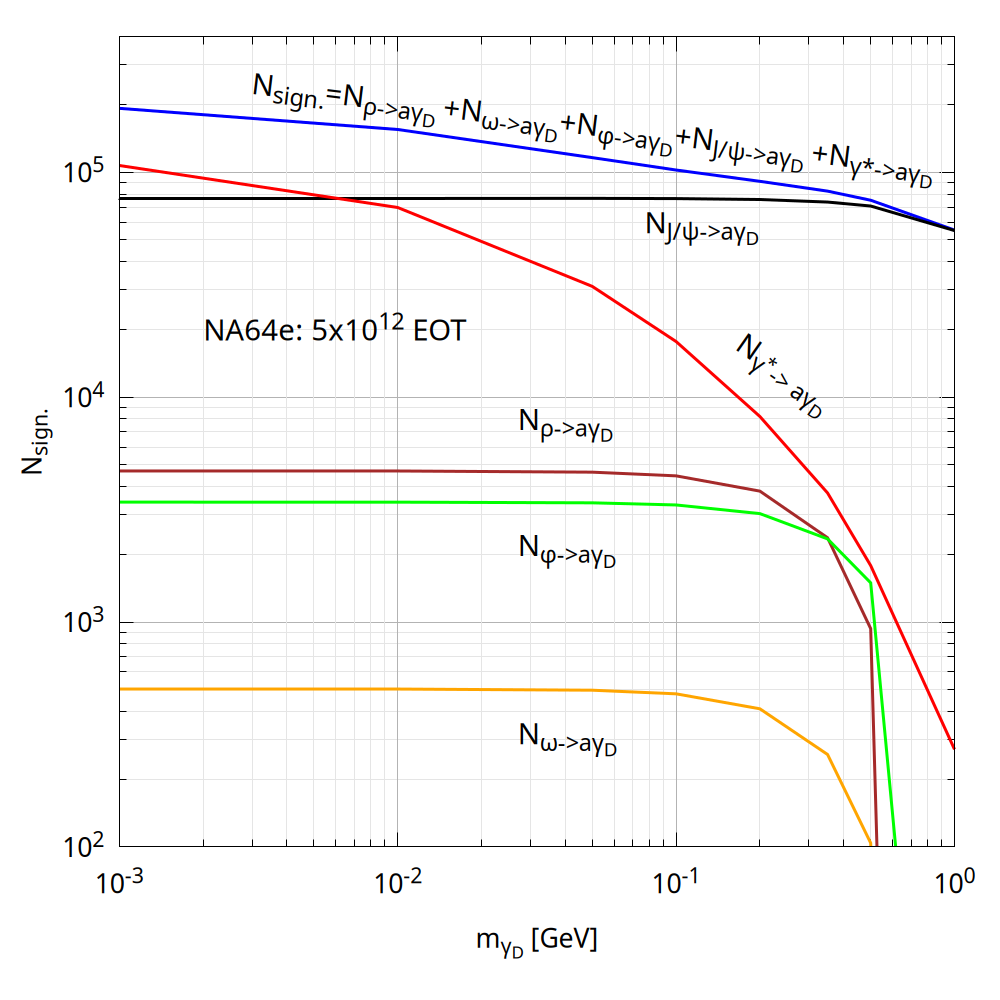}
\caption{The plot shows the  number of $a \gamma_D$ pairs produced versus $\gamma_D$ mass for both LDMX ($10^{16}$~EOT) and NA64e ($5\times 10^{12}$~EOT), with the coupling constant chosen to be $g_{a\gamma\gamma_D}=1~\mbox{GeV}^{-1}$. The hidden particles  are produced in bremsstrahlung-like reactions and various meson decays. {\it Left panel:} the resulted number of invisible signal events for the LDMX experiment, the $a\gamma_D$ pairs are produced through the bremsstrahlung-like channel $\gamma^*\to a\gamma_D$ (red line), rho meson  
$\rho \to a\gamma_D$ (brown  line), omega meson $\omega \to a\gamma_D$ (orange line), and phi meson decay $\phi \to a\gamma_D$ (green line). The resulted number of $a\gamma_D$ pairs produced at the LDMX is  shown (blue  line). {\it Right panel:} the same as in the left panel, but for the NA64e electron missing momentum facility, implying that  the dominant signal mechanisms are due to the  virtual photon emission $\gamma^* \to a\gamma_D$ (red  line) and $J/\psi\to a \gamma_D$ decays (black line). 
\label{NtotFig} }
\end{figure*}

The resulted number of missing energy events reads 
\begin{equation}
N_{\rm sign.} \simeq N_{\gamma^*\to a \gamma_D} + \sum_{V } N_{ V \to a \gamma_D }. 
\label{TotSignLDMX}
\end{equation}
In Fig.~\ref{NtotFig} we show the total number of $a\gamma_D$ pairs produced at the NA64e and LDMX experiments as a function of~$m_{\gamma_D}$. 

For the LDMX, the dominant production mechanism of relatively light dark photons, $m_{\gamma_D} \lesssim 30~\mbox{MeV}$, 
is associated with off shell photon emission, $\gamma^* \to a \gamma_D$, while for the heavier dark photons, $m_{\gamma_D } \gtrsim 100~\mbox{MeV}$, the relevant rate is comparable 
with vector meson one $V\to a \gamma_D$, i.~e.~the vector mesons provide a sizable contribution to the missing energy signal. Accounting for the $V$ meson contribution, would result in  
sensitivity enchantment of the LDMX to the dark ALP portal coupling, $g_{a\gamma\gamma_D}$, in the sub-GeV mass range,~$m_{\gamma_D} \lesssim 1~\mbox{GeV}$. 

The NA64e signal associated with coherent photoproduction of the $J/\psi$ meson yields a rate comparable to that of 
$\gamma^*\to a\gamma_D$ for sufficiently light hidden vectors, i.~e.~, for masses $m_{\gamma_D} \lesssim 10~\text{MeV}$. 
The total number of missing energy signatures depends weakly on dark photon mass $m_{\gamma_D}$, This implies a sufficiently mild slope of the graph for the sub-GeV dark photons.
The production of $\rho, \omega$ and $\phi$ vector mesons at  NA64$e$ is subdominant for the entire mass range of interest $1~\mbox{MeV} \lesssim m_{\gamma_D} \lesssim 1~\mbox{GeV}$. 

For both experiments, the yields of $\phi$ mesons are lower than the yields of $\omega$ mesons. Nevertheless, as shown in Fig.~\ref{NtotFig}, the invisible decays of $\phi$ mesons can provide better sensitivity due to the larger branching ratio compared to that of the $\omega$ meson.

\begin{table}
\caption{Parameteres of vector mesons (mass $m_V$, coupling constant $g_V$, total decay width $\Gamma_{\rm tot}^{V}$) relevant for our analysis.} 
\def\arraystretch{1.5}
\vspace*{.2cm}
\begin{tabular}{ l c c c }
    \hline
    \hline
    $V$ \ & \ $m_V$ (MeV) \ & \ $g_V$ \  
    & \ $\Gamma_{\rm tot}^{V}$ (MeV) \\ 
      \hline
      \hline
    $\rho$     & 775   & 4.96  & 147     \\   
    $\omega$   & 782   & 17.3  & 8.6     \\
    $\phi$     & 1019  & 13.4  & 4.2     \\
    $J/\psi$   & 3097  & 11.8  & 0.093   \\
    $\Upsilon$ & 9460  & 40.3  & 0.054   \\
    \hline
    \hline
\end{tabular}
\label{tab:VecMeSTypTerms}
\end{table}

\section{Benchmark experiments
\label{Sec:Experimenta}}

The missing energy analysis for the process \( eN \to eN a\gamma_D \) applied to both NA64e and LDMX builds upon the established frameworks developed for the related dark photon search \( eN \to eN\gamma_D \), as detailed in 
Refs.~\cite{NA64:2023wbi} and \cite{LDMX:2025bog}, respectively. The experimental signatures for these two channels are nearly identical: both experiments search for events with a single scattered electron in the final state and a large 
missing energy, typically exceeding \( \gtrsim 50\% \) of the initial beam energy. This stringent requirement on the missing energy fraction, enabled by a fully hermetic detector design, provides a powerful tool for suppressing standard 
model backgrounds.

We now outline the benchmark input parameters used for each experimental setup.

\subsection{The NA64$e$ experiment} 

The NA64$e$ employs  the optimized
electron beam from the H4 beam line at the SPS at CERN. 
The  maximum intensity of the beam is $\simeq 10^7$ electrons per spill of 
$4.8\,\mbox{s}$, the number of good spills is estimated to be $4000$  per day.

The reaction  for dark photons and ALPs production, $e N \to e N a \gamma_D$,  involves the scattering of a high-energy electron beam (\(E_e = 100\) GeV) off heavy nuclei of a lead-scintillator electromagnetic calorimeter (ECAL) surving as an active target,  followed by the prompt invisible decay 
\(\gamma_D \to \bar{\chi}\chi\). A significant fraction of the primary beam energy, \(E_{\rm miss}\), is carried away by the \(a\gamma_D\) system (and subsequently by the dark sector fermions), leaving no detectable signal in the hermetic NA64\(e\) detector. The remaining energy, \(E_e^{\rm rec} \equiv E_e - E_{\rm miss}\), is deposited in the ECAL by the scattered electron. Consequently, the production of hidden particles manifests as an excess of events containing a single electromagnetic shower with energy \(E_e^{\rm rec}\) above the expected background (see, e.g., Ref.~\cite{NA64:2016oww}). The ECAL target of NA64e is a matrix of $6\times 6$ 
Shashlyk-type modules assembled from lead (Pb) 
$(\rho=11.34\,\mbox{g cm}^{-3}, A=207\,\mbox{g mole}^{-1}, Z=82)$ 
and scintillator (Sc) plates.  Note that production of hidden particles in the scintillator 
is subleading due to its larger radiation length, $X_0(Sc)\gg X_0(Pb)$, thus we ignore it in the  calculation.  

Additionally, we  assume that the electromagnetic shower in the ECAL is initiated  within its  typical radiation length of $0.56~\mbox{cm}$. This defines the effective interaction length in Eq.~(\ref{NumberOfMissingEv1}) as \(L_T \simeq X_0\), implying that the dominant production of hidden particles occurs within the initial radiation length of the active ECAL~\cite{Chu:2018qrm}. Candidate events are required to satisfy \(E_e^{\rm rec} \lesssim 0.5 E_e\), corresponding to a lower integration limit of \(E_{\rm min} = 50\) GeV in Eq.~(\ref{NumberOfMissingEv1}).

The dominant background sources for the NA64\(e\) missing-energy search are identified as \cite{NA64:2023wbi}: (i) muon pair losses or decays in the target; (ii) in-flight decays along the beamline; (iii) insufficient calorimeter hermeticity; and (iv) particles passing through the calorimeter without interaction. 

We adopt the background study presented in Ref.~\cite{NA64:2023wbi}, which reports an estimated background of \(b  \ll 1 \) events for the currently accumulated exposure of \(\mbox{EOT} \simeq 9.37 \times 10^{11}\). For the projected future exposure of \(\mbox{EOT} \lesssim  10^{13}\), we anticipate that upgraded detector electronics and refined veto systems will improve background rejection by an \(\mathcal{O}(1)\) factor. This improvement is expected to maintain the total background at a level \(b \lesssim 1\) event, ensuring a statistically negligible background for the planned sensitivity studies  with $\simeq  10^{13}$ EOT.

\subsection{The LDMX experiment} 
 
The LDMX is a proposed electron fixed-target experiment at Fermilab, sharing a similar beam-based approach with 
NA64e. LDMX is specifically designed to reconstruct the missing  momentum of the scattered electron, providing a complementary probe of the process in Eq.~(\ref{generalMissinEnergyProcess}). The stringent missing-momentum requirements and extensive active veto systems employed by both experiments render them virtually background-free for the signal topology under consideration~\cite{NA64:2016oww,Berlin:2018bsc}.

The conceptual design of LDMX comprises a thin target, a precision silicon tracker, and downstream 
electromagnetic and hadronic calorimeters. For detailed descriptions of the detector layout and subsystems, 
see Refs.~\cite{Berlin:2018bsc,Akesson:2022vza}.

A significant fraction of the primary electron beam energy is lost via the emission of dark-sector particles 
within a thin upstream target. The resulting missing momentum of the scattered electron is reconstructed using a 
downstream silicon tracker system complemented by electromagnetic and hadronic calorimeters.

For the numerical analysis presented here, we consider an aluminum target with the following properties: 
radiation length \(X_0 = 8.9~\text{cm}\), density \(\rho = 2.7~\text{g cm}^{-3}\), atomic mass 
\(A = 27~\text{g mol}^{-1}\), and atomic number \(Z = 13\). 
The effective target thickness is taken as \(L_T \simeq 0.4 X_0\). According to its design parameters, LDMX 
plans to accumulate an integrated luminosity of \(\mbox{EOT} \simeq 10^{16}\) electrons on target, with a beam 
energy of up to \(E_e = 8~\text{GeV}\) during its second operational phase \cite{Schuster:2021mlr}.

For the LDMX analysis, we adopt selection criteria based on prior studies searching for invisibly decaying dark 
photons in the sub-GeV mass range \cite{Schuster:2021mlr,Akesson:2022vza,Berlin:2018bsc}. 
This approach is well justified, as the experimental signatures for the processes \(eN \to eN a\gamma_D\) and \(eN \to eN \gamma_D\) are nearly identical. The similarity arises because the differential energy spectra of the \(a\gamma_D\) system 
and a single dark photon share a common kinematic shape, with production dominated by the forward scattering 
region. Consequently, we apply the same missing-energy cut used in those references, requiring the reconstructed 
electron energy to satisfy \(E_e^{\rm rec} \lesssim 0.3 E_e\).
This corresponds to a lower integration limit of \(E_{\rm min} = 0.7 E_e \simeq 5.6\) GeV in~Eq.~(\ref{NumberOfMissingEv1}) 
for an $E_{e}\simeq 8~\mbox{GeV}$ beam.

Based on current simulation studies, the residual backgrounds for LDMX originate primarily from four categories 
\cite{LDMX:2025bog}: (i) unbiased electron interactions, (ii) photo-nuclear reactions, (iii) electro-nuclear 
interactions, and (iv) muon conversion events. To maintain sensitivity at the planned exposure of 
\(\mbox{EOT} \simeq 10^{16}\), a targeted upgrade of the front-end electronics and trigger systems is 
underway \cite{LDMX:2025bog}. The objective of these detector improvements is to achieve an effectively 
background-free experiment, ensuring that the projected background remains below a single event across the light 
dark state parameter space targeted by the benchmark search \cite{Berlin:2018bsc,Akesson:2022vza}.

   \begin{figure}[t!]
\centering
\includegraphics[width=0.5\textwidth]{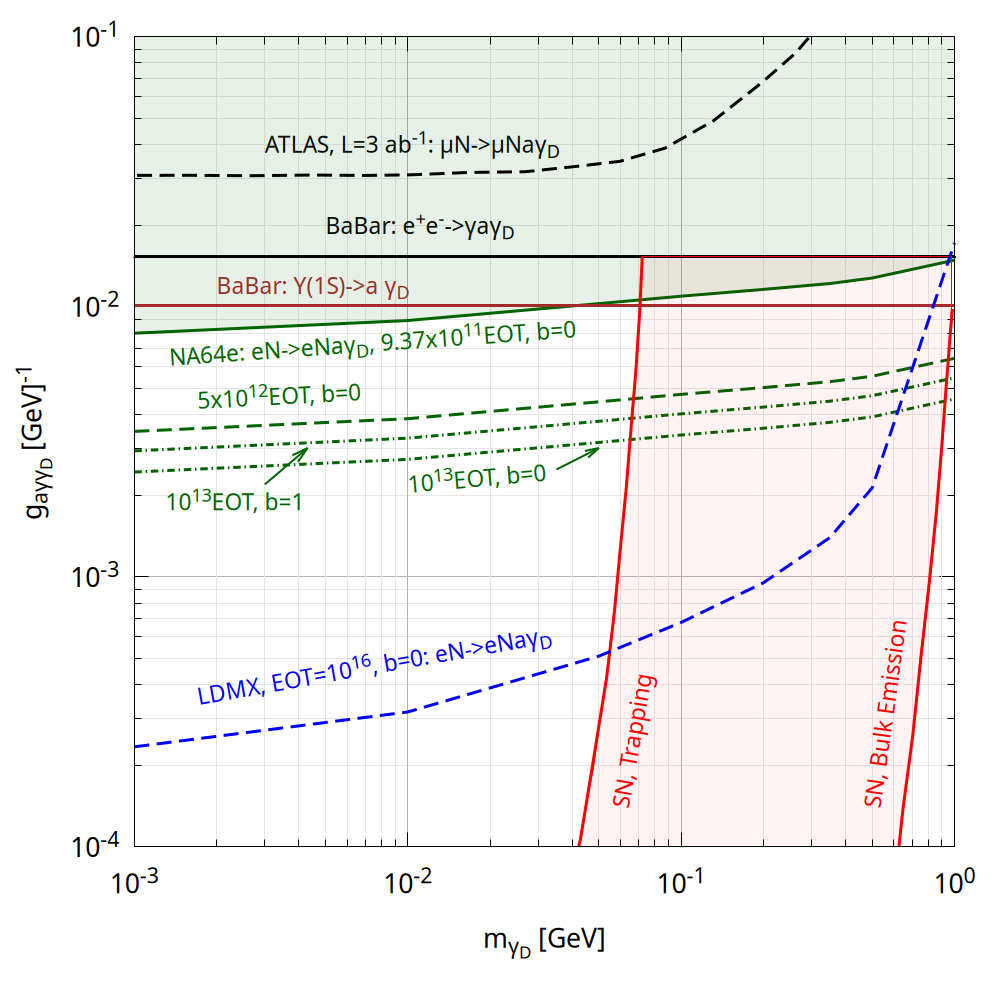}
\caption{The limits at $90~\%~\mbox{CL}$ on $g_{a\gamma\gamma_D}$ coupling from ATLAS, BaBar, and  the fixed-target experiments for the minimal dark
ALP portal setup as a function of  the dark photon mass $m_{\gamma_D}$. For the accelerator based sensitivity curves we 
imply that $\mbox{Br}(\gamma_D\to \chi \bar{\chi})\simeq 1$ and $m_a =10\, \mbox{keV}$.
Green dashed line is the  projected sensitivity for NA64$e$ experiment for $\mbox{EOT}\simeq 5\times 10^{12}$,  and blue  dashed line corresponds to the projected 
sensitivity of LDMX   facility for $\mbox{EOT}\simeq 10^{16}$.
The shaded green region shows the parameter space  excluded   by the NA64$e$  
experiment for $\mbox{EOT}\simeq 9.37\times 10^{11}$. For completeness, by  dashed dotted green lines  we  show the expected reaches of NA64e for $\mbox{EOT}\simeq 10^{13}$ with a finite background, $b\simeq 1$, and background free, $b\simeq 0$, case. 
We also show the BaBar~\cite{Zhevlakov:2022vio} excluded bounds for the data on monophoton
missing energy signature, $e^+e^-\to \gamma a \gamma_D$. The BaBar limits~\cite{BaBar:2009gco} from invisible decays 
$\Upsilon(1S) \to a\gamma_D $ are shown by brown solid line. The projected bounds of ATLAS 
($\mathcal{L}\simeq 3~\mbox{ab}^{-1}$)
associated with muon missing momentum process, $\mu N \to \mu N a \gamma_D$ are shown by black dashed line~\cite{Galon:2019owl,ATLAS:2016lqx}. The region of parameter space excluded by supernova (SN) is shown by shaded red area~\cite{Hook:2021ous}.  
\label{MinimalPortalFigExpectedReach} }
\end{figure} 

\section{Existed and expected sensitivities
\label{Sec:Sensitivity}}

Recent theoretical and phenomenological developments in the dark axion portal framework are summarized as follows. 
The regarding portal itself was formalized in Ref.~\cite{Gutierrez:2021gol}, which proposed that the coupled system of axions and dark photons could provide a viable mechanism to account for the observed dark matter relic density~\cite{Arias:2025nub}.

Subsequently, detailed phenomenological studies have explored 
its experimental signatures. 
Ref.~\cite{deNiverville:2018hrc} provided a comprehensive analysis of potential signals at B-factories, fixed-target 
neutrino experiments, and beam-dump facilities. Further investigations have focused on specific channels: the 
monophoton signature arising from the decay $\gamma_D \to a \gamma$ was examined for the experiments such as   LUXE–NPOD~\cite{Bai:2021gbm},  
SHIP~\cite{Alekhin:2015byh},  FASER~\cite{Feng:2022inv}, and LHC~\cite{ATLAS:2022vhr} in Refs.~\cite{deNiverville:2019xsx,Ness:2025klj,Jodlowski:2023yne,Arias:2025tvd}. While a complementary 
analysis for reactor neutrino experiments was presented in Ref.~\cite{Deniverville:2020rbv}. In addition, authors of  
 Ref.~\cite{Domcke:2021yuz} proposed a scenario that connects the dark axion portal phenomenology to a 
potential solution for the electroweak hierarchy problem.

As emphasized previously, our analysis focuses on the scenario where the dark photon decays invisibly into dark sector fermions, 
$\gamma_D \to \chi \bar{\chi}$ and $m_{\gamma_D}>m_a$. Consequently, the exclusion limits derived from searches for visible dark photon signatures are not directly applicable to 
the parameter space considered in the following subsections. 

\subsection{Collider based constraints}

An authors of Ref.~\cite{Galon:2019owl} proposed to utilize the ATLAS as a muon fixed-target experiment. To be more specific,
authors  considered a novel analysis for the ATLAS detector, which takes advantage of the two independent muon momentum measurements by the inner detector and the muon system.
They also provide the expected reach for the muon-philic sub-GeV hidden vector, 
$\mathcal{L}\supset g_{\gamma_D} A_\nu' \bar{\mu} \gamma_\nu \mu$  (see e.~g.~ left panel of Fig.~4 from 
Ref.~\cite{Galon:2019owl}) in $(m_{\gamma_D}, g_{\gamma_D})$ plane from
ATLAS experiment for projected luminosoty of $\mathcal{L}\simeq 3~\mbox{ab}^{-1}$, implying muon missing momentum signature, $\mu N \to \mu N \gamma_D$.
These limits can be relevant for our 
benchmark scenario, if one recasts the expected reach $(m_{\gamma_D}, g_{\gamma_D}) \to (m_{\gamma_D}, g_{a\gamma\gamma_D})$.  Specifically, one
can obtain the relevant for our analysis conservative limit on $g_{a\gamma\gamma_D}^{\rm ATLAS} $ from the plot
shown in Fig.~4 of Ref.~\cite{Galon:2019owl}.  For
 the mass range of   interest $1\, \mbox{MeV} \lesssim m_{\gamma_D} \lesssim 1\, \mbox{GeV}$ the rough estimate yields the following relation between couplings:
\begin{equation}
 g_{a\gamma\gamma_D}^{\rm ATLAS} \simeq  g_{\gamma_D}^{\rm ATLAS}\label{ATLASLim}
 \times g_{a\gamma\gamma_D}^{\rm NA64\mu}/g_{\gamma_D}^{\rm NA64\mu},
\end{equation}
where $g_{a\gamma\gamma_D}^{\rm NA64\mu}$ and 
$g_{\gamma_D}^{\rm NA64\mu}$ are the experimental reaches  of the muon missing momentum facility NA64$\mu$ adapted from Ref.~\cite{Galon:2019owl} and Ref.~\cite{Zhevlakov:2022vio}, respectively for the equivalent anticipated statistics of muons accumulated on target.  The conservative estimate (\ref{ATLASLim}) is justified by the  the similar missing energy  cuts and typical muon beam energies at ATLAS and NA64$\mu$. 

In Fig.~\ref{MinimalPortalFigExpectedReach}  we also show The BaBar excluded bounds for the data on monophoton
missing energy signature, $e^+e^-\to \gamma a \gamma_D$, adapted from~\cite{Zhevlakov:2022vio}. These BaBar bounds at the level of $g_{a\gamma\gamma_D} \lesssim 1.5\times 10^{-2}~\mbox{GeV}^{-1}$ rule out 
the projected limits of ATLAS, $g_{a\gamma\gamma_D}\lesssim 3 \times 10^{-2}~\mbox{GeV}^{-1}$,
which can operate as muon missing momentum facility~\cite{Galon:2019owl}. 

For the  sub-GeV dark photon masses, a sufficiently strong bound,
\begin{equation}
g_{a\gamma\gamma_D}\lesssim 2.0\times 10^{-2}~\mbox{GeV}^{-1},
\label{UpsilonBound}
\end{equation} can be set from BaBar 
data~\cite{BaBar:2009gco} on invisible $\Upsilon$ meson decay. Namely, the regarding branching fraction is constrained  to be 
\begin{equation}
\mbox{Br}_{\Upsilon\to {\rm inv.}}\lesssim 3\times 10^{-4}
\label{UpsilinBranching}
\end{equation}
at~$90\%~\mbox{CL}$. That limit can be associate with a
 relevant for our analysis bound on the invisible branching fraction of $\Upsilon(1S)\to a \gamma_D$ vector meson.  
In order to extract the coupling constant bound from this 
limit, we employ again VMD model  for $\Upsilon(1S)$ meson, implying that 
typical mixing with photons is estimated to be, $g_{\Upsilon}\simeq 40.3$ from 
its radiative decay into $e^+e^-$ pair. Specifically, the VMD constant can be linked with leptonic decay width, $\Gamma_{\Upsilon\to e^+e^-}\simeq 1.3~\mbox{keV}$ as follows, 
\begin{equation}
g_{\Upsilon} \simeq \left( \frac{4\pi \alpha^2 m_{\Upsilon}}{3 \Gamma_{\Upsilon \to e^+e^-}}  \right)^{1/2}\simeq 40.3,
\end{equation}
where $\alpha\simeq 1/137$ is a fine structure coupling constant, $m_{\Upsilon}\simeq 9460~\mbox{MeV}$ is a mass of $\Upsilon(1S)$ meson. The resulted bound (\ref{UpsilonBound}) is obtained by incorporating data in 
Tab.~\ref{tab:VecMeSTypTerms}, Eqs.~(\ref{UpsilinBranching}) and (\ref{VinvisDecW}).
The $\Upsilon$ invisible decay bounds (\ref{BarZeeALPPortal1}) rule out both BaBar monophoton, $e^+ e^-\to \gamma a \gamma_D$, limits and ATLAS muon missing energy, $\mu N \to \mu N a \gamma_D$, projected sensitivities. 

\subsection{Astrophysical constraints}

In this subsection we follow Ref.~\cite{Hook:2021ous} to  overview briefly other constrains on dark axion portal coupling (\ref{LDAP}) coming from the supernova bounds.

The supernova SN1987A has served as an important astrophysical laboratory for probing physics beyond the SM. The impact of novel particle emission on supernova cooling is characterized by two distinct transport regimes, defined by the  interaction strength of the new particle species. In the weak-coupling limit, particles produced in the stellar core undergo negligible re-scattering, freely escaping the proto-neutron star volume. This defines the transparent or free-streaming regime, where energy 
loss is proportional to the production rate. Constraints derived in this regime establish a lower bound on the particle coupling, as any stronger interaction would only enhance the emission and violate observational limits.

Conversely, if the coupling is sufficiently strong, the novel particles achieve local thermal equilibrium and become trapped within the dense medium. In this trapping regime, energy is transported diffusively and radiated from a neutrinosphere-
like last scattering surface. As the coupling strength increases, this effective emission surface recedes to larger radii and lower temperatures within the stellar gradient, thereby reducing the spectral luminosity. This behavior generates an upper 
bound on allowable couplings, as excessively strong interactions suppress the emergent flux below levels detectable from SN1987A. The full exclusion contour, spanning from the 
free-streaming lower bound to the trapping upper bound, is plotted within the framework of the Raffelt criterion~\cite{Raffelt:1996wa}. In Fig.~\ref{MinimalPortalFigExpectedReach}
we show the relevant bounds by shaded red region, adapted from~\cite{Hook:2021ous}.

\subsection{The reach of the missing energy experiments}

By using Eq.~(\ref{TotSignLDMX}) to calculate the expected number of \(a\gamma_D\) production events, we derive projected exclusion limits on the portal coupling \(g_{a\gamma\gamma_D}\) within the minimal dark ALP portal framework. Assuming a 
background-free search ($b\simeq 0$) and a null observation, $n_{\rm obs.}\simeq 0$, we set a 90\% confidence level (CL) 
exclusion criterion corresponding to a signal yield of \(N_{\rm sign.} \gtrsim 2.3\) 
events. This criterion is used to determine the upper bounds on 
\(g_{a\gamma\gamma_D}\) presented in the following analysis.

In Fig.~\ref{MinimalPortalFigExpectedReach} we show the expected reach of NA64$e$ and LDMX by green and blue solid lines, respectively. Note that  projected  limits on $g_{a\gamma\gamma_D}$ from LDMX are fairly strong, i.~e.~for small masses of dark photon, $m_{\gamma_D}\simeq \mathcal{O}(1)~\mbox{MeV}$ the expected reach is estimated to be at the level of $g\lesssim \mathcal{O}(10^{-4})~\mbox{GeV}^{-1}$. 
The regarding LDMX sensitivity enchantment  can be explained by 
the large number of the projected accumulated statistics, $\mbox{EOT}\simeq 10^{16}$, by the final phase 
of  experimental  running.

\begin{table*}
\caption{Upper limits on couplings of new particles 
from data on EDMs of electron, muon, and neutron.} 
\def\arraystretch{1.5}
\begin{tabular}{c c c c}
\hline
\hline
 \ \ Coupling combination   \ \ & \ \ Electron  ($90~\%$ CL) \ \ & \ \ Muon  ($95~\%$ CL) \ \ & \ \ Neutron  ($90~\%$ CL) \ \
 \\
\hline
\hline
$m_{\gamma_D}=m_a \gg m_f$,  $ |g_f^a \, g_{a\gamma\gamma_D} e \epsilon_f |\times \left[\log\left(\Lambda^2/m^2_{\gamma_D}\right)+1/2\right]$: & &  & 
\\ 
  & $\lesssim 5.0 \times 10^{-15}~\mbox{GeV}^{-1},$  & $ \lesssim 2.3 \times 10^{-4}~\mbox{GeV}^{-1},$  & $ \lesssim 2.2 \times 10^{-11}~\mbox{GeV}^{-1},$ 
\\
 $m_{\gamma_D}\gg m_f \gg m_a$,  $ |g_f^a \, g_{a\gamma\gamma_D} e \epsilon_f |\times \left[\log\left(\Lambda^2/m^2_{\gamma_D}\right)+5/2\right]$: & &  & 
\\ 
\hline 
\hline
\end{tabular}
\label{tab:limits}
\end{table*}

In addition, we note that the projected limits of  NA64$e$ for $\mbox{EOT}\simeq 5\times 10^{12}$ can rule the typical couplings at the level of $g_{a\gamma\gamma_D} \lesssim  (3.5 - 5)\times 10^{-3}~\mbox{GeV}^{-1}$ for sub-GeV dark photons.   
In Fig.~\ref{MinimalPortalFigExpectedReach} we show by the shaded green region the excluded limits of NA64$e$ for the current accumulated  statistics~\cite{Andreev:2021fzd} 
of $\mbox{EOT}\simeq 9.37 \times 10^{11}$.  To be more specific, the current bounds of  NA64$e$ 
 rules out the  couplings in the range $8\times10^{-3} ~\mbox{GeV}^{-1} \lesssim  g_{a\gamma\gamma_D} \lesssim  1.5\times 10^{-2}~\mbox{GeV}^{-1}$ for the dark photon masses  of interest, $m_{\gamma_D} \lesssim 1~\mbox{GeV}$. 
 
 To perform additional conservative projection, in Fig.~\ref{MinimalPortalFigExpectedReach} we also show 
 the expected reaches of NA64e implying the anticipated statistics of $\mbox{EOT}\simeq 10^{13}$ for 
 both background free, $b=0$, and non-negligible number of regarding events, $b\simeq 1$ (see, e.~g., Ref.~\cite{Gninenko:2025aek} and references therein for detail).   
To summarize  principal findings of this section, we note that accounting for vector  mesons would result in a sensitivity enhancement by a factor of $\mathcal{O}(1)$ for the LDMX  facility in the large mass region, 
\( m_{\gamma_D} \lesssim 1~\mbox{GeV} \). In contrast, for the NA64e experiment,  vector meson decays enhance its  sensitivity to dark ALP portal by several orders of magnitude for the sub-GeV mass range. However, a sufficiently large portion of the parameter space within, $500~\mbox{MeV}~\lesssim  m_{\gamma_D} \lesssim 1~\mbox{GeV}$ is constrained from the~SN1987A. 

\begin{figure}[!t]
\centering
\includegraphics[width=0.5\textwidth]{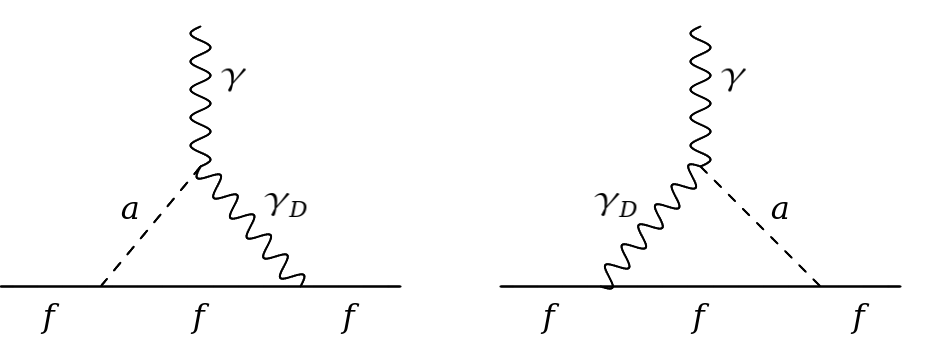}
\caption{Feynman diagrams which generate EDM terms due to axion coupling with both  dark and SM photons 
(see, e.~g., Eq.~(\ref{LagrangianVectorAxionEDM}) for detail).    \label{BarZeeALPPortal1}}
\end{figure}

\begin{figure}[t!]		
	\includegraphics[width=0.49\textwidth]{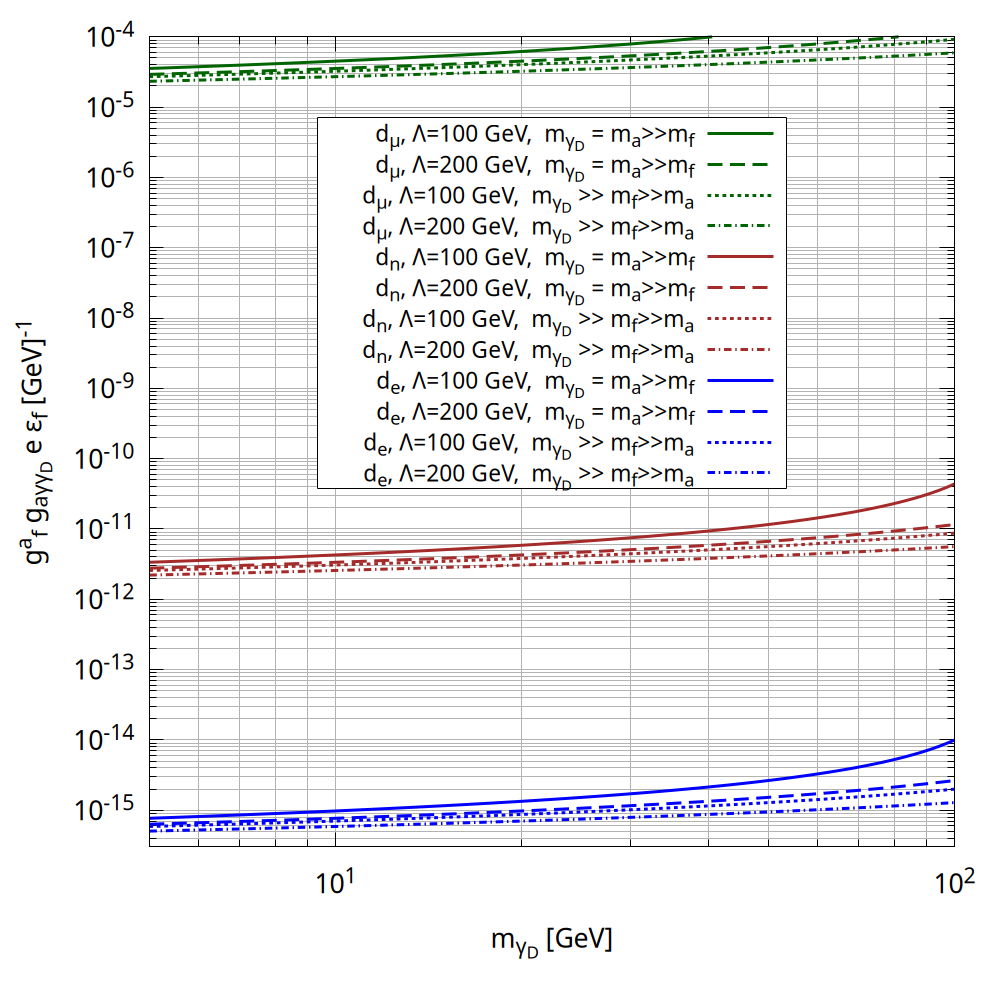}
	\caption{Boundaries for the product of CP-odd and CP-even couplings from EDM neutron and electron/muon as the function of the dark photon mass.} 
	\label{from_edm}
\end{figure}

\section{ Bounds from fermion EDMs
\label{EDMsectionLabel}}

In this section, we derive the constraints on the combinations
of $P$-even and $P$-odd couplings of new particles using
data on electric dipole moments (EDMs) of leptons and neutron.
The contributions of new particles to EDMs are described
by the Bar-Zee diagrams in Fig.~\ref{BarZeeALPPortal1}.
This contribution is generated by an additional interaction Lagrangian containing three terms: 
(i) $P$-parity violating coupling of ALP with SM fermions, $g_f^a$, 
(ii) $P$-parity conserving coupling of ALP with SM photon and dark photon, $g_{a\gamma\gamma_D}$, 
and (iii) $P$-parity conserving coupling of dark photon with SM fermions, $e \epsilon_f$, 
\begin{eqnarray}
\mathcal{L} \supset   g^{a}_f \, a \bar{f} f + \frac{g_{a \gamma \gamma_D}}{2} a F_{\mu\nu} \widetilde{F}^{\prime\mu\nu} 
 + e \epsilon_f \bar{f} \gamma^\mu A_\mu' f \,.  
\label{LagrangianVectorAxionEDM}
\end{eqnarray}

To be more specific, the EDM of spin-$\frac{1}{2}$ fermion $f$ (neutron or leptons) 
is defined as $d_f=D_E(0,\Lambda)$, where $D_E(q^2,\Lambda)$ is the relativistic electric dipole form factor extracted from full electromagnetic vertex function of corresponding fermion~\cite{Gutsche:2016jap,Zhevlakov:2019ymi,Zhevlakov:2018rwo,Dib:2006hk,Faessler:2006vi}: 
\eq 
M_{\rm inv} &=& \bar u_f(p_2)\,\Gamma^\mu(p_1,p_2)\,u_f(p_1)\,,
\nonumber\\
\Gamma^\mu &=& 
- \sigma^{\mu\nu}q_{\nu} \gamma^5 \, D_E(q^2,\Lambda) 
\,+\, \ldots, 
\label{vertex}
\en
where $q = p_1-p_2$ is a momentum of photon. The contribution of dark axion portal depicted in Fig.~\ref{BarZeeALPPortal1} is
divergent. With the cut-off regularization, the result in (\ref{vertex}) is expressed in terms of the ultra-violet (UV) scale,~$\Lambda$. 

In particular, the relevant contributions of the specific diagrams in Fig.~\ref{BarZeeALPPortal1} read~\cite{Zhevlakov:2022vio}
\eq
d_f = - \frac{G}{4 \pi^2} \, \biggl[ - \frac{1}{2} + 
\frac{I(m_{\gamma_D}^2, m_f^2) - I(m_{a}^2, m_f^2)}{m_{\gamma_D}^2 - m_{a}^2} \biggr] 
\,, 
\en
where  $G = g^a_f g_{a\gamma\gamma_D} e \epsilon_f $ is a product of the coupling constants, and auxiliary loop function 
$I(m^2, m_f^2)$ is defined as 
\eq
I(m^2, m_f^2) &=& \int\limits_0^1 dx \, 
\Big[m^2 (1-x) + m_f^2 x^2\Big]  \nonumber\\ 
&\times& \log\frac{m^2 (1-x) + m_f^2 x^2}{\Lambda^2}. 
\en 
Since the  interaction in (\ref{LDAP}) is non-renormalizable, we treat $\Lambda$ as scale up to which our consideration is valid. 
Using the upper limits/results for the electron, muon, and neutron EDMs: 
\eq 
& &|d_e| < 4.1 \times 10^{-30} \, e \, \mbox{cm, \quad at \,\, 90 \%\, CL,}\,\, 
\mbox{Ref.~\cite{Roussy:2022cmp}}\,,  
\nonumber\\
& &|d_\mu| < 1.9  \times 10^{-19} \, e \, \mbox{cm, \quad at \,\, 95 \%\, CL,}\,\, 
\mbox{Ref.~\cite{Muong-2:2008ebm}}\,, 
\nonumber\\
& &|d_n| < 1.8 \times 10^{-26} \, e \, \mbox{cm, \quad at \,\, 90 \%\, CL,}\,\, 
\mbox{Ref.~\cite{Abel:2020pzs}} \,,  
\nonumber
\en 
we get the  upper limits for combinations of couplings of 
new particles, which are displayed in Tab.~\ref{tab:limits}. 
We set for concreteness two 
benchmark cut-off scales $\Lambda\simeq 100~\mbox{GeV}$ and  $\Lambda\simeq 200~\mbox{GeV}$ 
and plot relevant limits in  Fig.~\ref{from_edm} for the following approaches: 
    (i) heavy dark photon with mass hierarchy, 
    $m_{\gamma_D} \gg m_a \gg m_f$, 
    (ii) equivalent mass of dark photon and ALP, 
    $m_{\gamma_D} =  m_a \gg m_f$.
    
It turns out that the electron EDM, which is currently constrained
by the JILA collaboration~\cite{Roussy:2022cmp} is very sensitive to probe non-minimal CP-odd dark ALP portal scenario (\ref{LagrangianVectorAxionEDM}) at the 
level of $|g_e^a g_{a\gamma\gamma_D} e \epsilon_e| \lesssim \mathcal{O}(10^{-15})~\mbox{GeV}^{-1}$.
The upper bounds on the coupling combination for the muon - and neutron- specific scenarios 
are estimated to be at the level of $|g_\mu^a g_{a\gamma\gamma_D} e \epsilon_\mu| \lesssim \mathcal{O}(10^{-5})~\mbox{GeV}^{-1}$ and $|g_n^a g_{a\gamma\gamma_D} e \epsilon_n| \lesssim \mathcal{O}(10^{-12})~\mbox{GeV}^{-1}$, respectively. The bounds on the combinations of couplings depend logarithmically on dark photom mass

\section{Conclusion
\label{ResultsSection}}

We have studied  the missing energy signatures for the projected and existed electron  fixed target experiments, such as LDMX and  NA64$e$. In particular, we calculated the rate of $a \gamma_D$ pair production processes  $e N \to e N a \gamma_D$ followed by the invisible dark photon decay into DS  particles  $\gamma_D\to \chi \bar{\chi}$ for the specific fixed target facility.   

We derive new exclusion limits based on the NA64\(e\) data set with \(9.37 \times 10^{11}~\mbox{EOT}\)  by considering two distinct production mechanisms leading to invisible final states: (i) the bremsstrahlung-like emission of an ALP-dark photon pair, \( e N \to e N \gamma^* (\to a \gamma_D) \), and (ii) exclusive vector meson photoproduction, \( \gamma^* N \to N V \), followed by the invisible decay \( V \to a \gamma_D \).

We find that vector mesons enhance sensitivity differently across experiments. For LDMX in the \( m_{\gamma_D} \lesssim 1~\text{GeV} \) region, the enhancement factor is $\mathcal{O}(1)$. For NA64, vector meson decays boost sensitivity to the dark ALP portal by several orders of magnitude.

In a separate analysis, we establish constraints on the parameter space of \(CP\)-violating, fermion-specific ALP couplings. 
These bounds are obtained by correlating the contributions of 
such couplings to the electric dipole moments of the SM fermions. Specifically, we utilize current experimental limits 
on the  EDM of SM fermions and include the associated, 
loop-induced contributions to the EDMs of the electron, muon, 
and neutron.

\begin{acknowledgments} 
We would like to thank I.~Galon, S.~Demidov, R.~Dusaev, A.~Pukhov, and Y.~Soreq, for very helpful discussions and  
correspondences. The  work of V.~E.~L. and S.~K. on deriving EDM bounds was funded by FONDECYT (Chile) under Grant 
No. 1240066 and by ANID$-$Millen\-nium Program$-$ICN2019\_044 (Chile).
The work of D.~V.~K on deriving sensitivities for both NA64$e$ and ATLAS experiments was  
supported by the the Russian Science Foundation grant No.~25-12-00309.

\end{acknowledgments}	

\bibliography{bibl}

\end{document}